\documentclass[preprint,12pt]{article}

\usepackage{amsmath,amssymb,array,calc,epsfig,psfrag, amscd, datetime}
\usepackage{rotating}

\usepackage{color}
\usepackage[
      colorlinks=true,
      urlcolor=blue,    
      filecolor=blue,     
      citecolor=red,
      pdfstartview=FitV,
       bookmarksopen=true    
      ]{hyperref}
\usepackage[left=3cm,top=2.5cm,right=3cm,nohead]{geometry}

\newcommand{\nc}{\newcommand}

\definecolor{cardinal}{rgb}{0.6,0,0}
\definecolor{darkgreen}{rgb}{0,0.5,0}
\definecolor{golden}{rgb}{0.92, 0.7, 0}
\definecolor{midnight}{rgb}{0, 0, 0.5}
\definecolor{darkblue}{rgb}{0.2, 0, 0.8}

\nc{\ra}{\rightarrow} 
\nc{\lra}{\leftrightarrow} 
\nc{\Ra}{\Rightarrow} 
\nc{\LRa}{\Leftightarrow} 
\nc{\blp}{{\big (}}
\nc{\brp}{{\big )}}
\nc{\Blp}{{\Big (}}
\nc{\Brp}{{\Big )}}
\nc{\bglp}{{\bigg (}}
\nc{\bgrp}{{\bigg )}}
\nc{\Bglp}{{\Bigg (}}
\nc{\Bgrp}{{\Bigg )}}
\nc{\slb}{{\rm [}}
\nc{\srb}{{\rm ]}}
\nc{\bslb}{{\rm \big [}}
\nc{\bsrb}{{\rm \big ]}}
\nc{\Bslb}{{\rm \Big [}}
\nc{\Bsrb}{{\rm \Big ]}}

\def\al{\alpha}

\def\eps{\epsilon}
\nc{\veps}{\varepsilon}
\def\gam{\gamma}

\def\lam{\lambda}
\def\om{\omega}

\nc{\vphi}{\varphi}
\def\tha{\theta}

\def\sig{\sigma}

\def\Gam{\Gamma}

\def\Lam{\Lambda}
\def\Om{\Omega}
\def\Sig{\Sigma}

\def\coeff#1#2{\relax{\textstyle {#1 \over #2}}\displaystyle}

\nc{\myvspace}{\rule[-1em]{0pt}{2.5em}}
\nc{\bea}{\begin{eqnarray}}
\nc{\eea}{\end{eqnarray}}
\nc{\be}{\begin{equation}}
\nc{\ee}{\end{equation}}
\nc{\barr}{\begin{array}}
\nc{\earr}{\end{array}}

\nc{\co}{{\cal o}}

\nc{\cA}{{\cal A}}
\nc{\cB}{ \cal B}

\def\cD{{\cal D}}

\nc{\cF}{{\cal F}}
\nc{\cG}{{\cal G}}

\def\cI{{\cal I}}

\def\cK{{\cal K}}
\nc{\cL}{{\cal L}}
\nc{\cM}{{\cal M}}
\def\N{{\cal N}}
\def\cN{{\cal N}}

\nc{\cQ}{{\cal Q}}
\nc{\cR}{{\cal R}}
\def\cS{{\cal S}}

\def\cV{{\cal V}}
\def\cV{{\cal V}}
\def\cW{{\cal W}}

\def\cZ{{\cal Z}}
\nc{\cQd}{\cQ^{\dagger}}
\nc{\cRd}{\cR^{\dagger}}
\nc{\BB}{{\mathbb B}}
\nc{\CC}{{\mathbb C}}
\nc{\DD}{{\mathbb D}}
\nc{\EE}{{\mathbb E}}
\nc{\FF}{{\mathbb F}}
\nc{\GG}{{\mathbb G}}
\nc{\HH}{{\mathbb H}}
\nc{\JJ}{{\mathbb J}}
\nc{\MM}{{\mathbb M}}
\nc{\RR}{{\mathbb R}}
\nc{\PP}{{\mathbb P}}
\nc{\QQ}{{\mathbb Q}}
\nc{\UU}{{\mathbb U}}
\nc{\ZZ}{{\mathbb Z}}
\nc{\calone}{{\mathbb 1}}

\nc{\half}{\coeff{1}{2}}
\nc{\quarter}{\coeff{1}{4}}
\nc{\del}{\partial}

\nc{\delbar}{\bar\partial}
\nc{\thalf}{\frac{t}{2}}
\nc{\Spin}{\operatorname{Spin}}
\nc{\SO}{\operatorname{SO}}

\nc{\Sp}{{\rm Sp}}
\nc{\com}[2]{{ \left[ #1, #2 \right] }}
\nc{\acom}[2]{{ \left\{ #1, #2 \right\} }}
\nc{\rr}{\rightarrow}
\nc{\p}{\partial}
\nc{\LT}{{\LL_\T}}
\nc{\Tr}{{\rm Tr}}
\nc{\tr}{{\rm tr}}
\nc{\Adag}{A^{\dagger}}
\nc{\AdagI}{A^{\dagger I}}
\nc{\AdagJ}{A^{\dagger J}}
\nc{\AdagK}{A^{\dagger K}}
\nc{\AdagL}{A^{\dagger L}}
\nc{\AdagM}{A^{\dagger M}}
\nc{\Bdag}{B^{\dagger}}
\nc{\BdagI}{B^{\dagger}_I}
\nc{\BdagJ}{B^{\dagger}_J}
\nc{\BdagK}{B^{\dagger}_K}
\nc{\BdagL}{B^{\dagger}_L}
\nc{\BdagM}{B^{\dagger}_M}
\nc{\Cdag}{C^{\dagger}}
\nc{\CdagI}{C^{\dagger I}}
\nc{\CdagJ}{C^{\dagger J}}
\nc{\CdagK}{C^{\dagger K}}
\nc{\Ddag}{D^{\dagger}}
\nc{\DdagI}{D^{\dagger I}}
\nc{\DdagJ}{D^{\dagger J}}
\nc{\DdagK}{D^{\dagger K}}
\nc{\ttha}{\tilde{\theta}}
\nc{\ttau}{\tilde{\tau}}
\nc{\tTha}{\tilde{\Theta}}
\nc{\tphi}{\tilde{\phi}}
\nc{\tsig}{\tilde{\sig}}
\nc{\tom}{\widetilde{\om}}
\nc{\tOm}{\widetilde{\Om}}
\nc{\tlam}{\widetilde{\lam}}
\nc{\tLam}{\tilde{\Lam}}
\nc{\tSig}{\widetilde{\Sig}}
\nc{\tPhi}{\tilde{\Phi}}
\nc{\tPhibar}{\ol{\tPhi}}
\nc{\tPi}{\widetilde{\Pi}}
\nc{\tpsi}{\widetilde{\psi}}
\nc{\tPsi}{\tilde{\Psi}}
\nc{\tgam}{\widetilde{\gam}}
\nc{\tGam}{\widetilde{\Gam}}
\nc{\tzeta}{\tilde{\zeta}}
\nc{\tZeta}{\tilde{\Zeta}}
\nc{\teta}{\widetilde{\eta}}
\nc{\teps}{\tilde{\eps}}
\nc{\tveps}{\tilde{\veps}}
\nc{\tEta}{\tilde{\Eta}}
\nc{\tchi}{\tilde{\chi}}
\nc{\tChi}{\tilde{\Chi}}
\nc{\txi}{\tilde{\xi}}
\nc{\tXi}{\widetilde{\Xi}}
\nc{\tnu}{\tilde{\nu}}
\nc{\tmu}{\tilde{\mu}}

\nc{\tb}{\tilde b}
\nc{\tc}{\tilde c}
\nc{\te}{\tilde e}
\nc{\tf}{\widetilde f}
\nc{\tg}{\widetilde g}
\nc{\ti}{\tilde i}
\nc{\tj}{\tilde j}
\nc{\tk}{\widetilde k}
\nc{\tl}{\tilde l}
\nc{\tm}{\widetilde m}
\nc{\tn}{\tilde n}
\nc{\tp}{\tilde{p}}
\nc{\tq}{\widetilde{q}}
\nc{\trr}{{\tilde r}}
\nc{\ts}{{\tilde s}}
\nc{\tu}{{\tilde u}}
\nc{\tv}{{\tilde v}}
\nc{\tw}{{\tilde w}}
\nc{\tx}{{\tilde x}}
\nc{\ty}{{\tilde y}}
\nc{\tz}{\tilde z}
\nc{\tA}{{\widetilde A}}
\nc{\tAbar}{{\ol \tA}}
\nc{\tB}{{\widetilde B}}
\nc{\tC}{{\widetilde C}}
\nc{\tD}{{\widetilde D}}
\nc{\tE}{{\widetilde E}}
\nc{\tF}{{\widetilde F}}
\nc{\tG}{{\widetilde G}}
\nc{\tcG}{{\widetilde \cG}}
\nc{\tH}{{\widetilde H}}
\nc{\tI}{{\widetilde I}}
\nc{\tcI}{{\widetilde \cI}}
\nc{\tJ}{{\widetilde J}}
\nc{\tJbar}{{\ol {\tilde J}}}
\nc{\tK}{{\widetilde K}}
\nc{\tL}{{\widetilde L}}
\nc{\tcL}{{\widetilde \cL}}
\nc{\tcLbar}{{\ol \tcL}}
\nc{\tM}{{\widetilde M}}
\nc{\tN}{{\widetilde N}}
\nc{\tcN}{{\widetilde \cN}}
\nc{\tP}{{\widetilde P}}
\nc{\tQ}{{\widetilde Q}}
\nc{\tR}{{\widetilde R}}
\nc{\tS}{\widetilde{S}}
\nc{\tT}{\widetilde{T}}
\nc{\tU}{\widetilde{U}}
\nc{\tUU}{\widetilde{\UU}}
\nc{\tV}{\widetilde{V}}
\nc{\tcV}{\widetilde{\cV}}
\nc{\tW}{\widetilde{W}}
\nc{\tcF}{\widetilde{{\cal F}}}
\nc{\tX}{\widetilde{X}}
\nc{\tY}{\widetilde{Y}}
\nc{\tcZ}{\tilde{\cZ}}
\nc{\tcZbar}{\ol{\tcZ}}

\nc{\ha}{\hat a}
\nc{\hb}{\hat b}
\nc{\hc}{\widehat c}
\nc{\hd}{\widehat d}
\nc{\he}{\widehat e}
\nc{\hf}{\widehat f}
\nc{\hg}{\widehat g}
\nc{\hh}{\widehat h}
\nc{\hm}{\widehat m}
\nc{\hn}{\widehat n}
\nc{\hp}{\widehat p}
\nc{\hq}{\widehat q}
\nc{\hr}{\widehat r}
\nc{\hs}{\widehat s}
\nc{\hv}{\widehat v}
\nc{\hw}{\widehat w}
\nc{\hx}{\widehat x}
\nc{\hy}{\widehat y}
\nc{\hz}{\widehat z}
\nc{\zhat}{\hat z}
\nc{\hA}{\widehat{A}}
\nc{\hB}{\widehat{B}}
\nc{\hC}{\widehat{C}}
\nc{\hD}{\widehat{D}}
\nc{\hE}{\widehat{E}}
\nc{\hF}{\widehat{F}}
\nc{\hcF}{\widehat{\cF}}
\nc{\hG}{\widehat{G}}
\nc{\hH}{\widehat{H}}
\nc{\hI}{\widehat{I}}
\nc{\hcI}{\widehat{\cI}}
\nc{\hJ}{\widehat{J}}
\nc{\hK}{\widehat{K}}
\nc{\hL}{\widehat{L}}
\nc{\hcL}{\widehat{\cL}}
\nc{\hM}{\widehat M}
\nc{\hcM}{\widehat{\cM}}
\nc{\hN}{\widehat{N}}
\nc{\hO}{\widehat{O}}
\nc{\hP}{\widehat{P}}
\nc{\hQ}{\widehat{Q}}
\nc{\hcR}{\widehat{\cR}}
\nc{\hR}{\widehat{R}}
\nc{\hS}{\widehat{S}}
\nc{\hcS}{\widehat{\cS}}
\nc{\hT}{\widehat{T}}
\nc{\hU}{\widehat{U}}
\nc{\hV}{\widehat V}
\nc{\hcV}{\widehat \cV}
\nc{\hX}{\widehat X}
\nc{\hcZ}{\widehat \cZ}
\nc{\hcZbar}{\ol{\widehat \cZ}}

\nc{\heta}{\widehat{\eta}}
\nc{\hal}{\widehat \alpha}
\nc{\hbeta}{\widehat \beta}
\nc{\hphi}{\widehat{\phi}}
\nc{\hkap}{\hat{\kappa}}
\nc{\hchi}{\widehat{\chi}}
\nc{\hpsi}{\widehat{\psi}}
\nc{\hgam}{\widehat{\gam}}
\nc{\hPhi}{\hat{\Phi}}
\nc{\hPsi}{\hat{\Psi}}
\nc{\hGam}{\hat{\Gam}}
\nc{\omhat}{\widehat{\om}}
\nc{\htha}{\hat{\tha}}
\nc{\hrho}{\widehat{\rho}}
\nc{\hdel}{\widehat{\del}}

\nc{\w}{\wedge}

\nc{\vb}{\vec b}
\nc{\vc}{\vec c}
\nc{\vd}{\vec d}
\nc{\ve}{\vec e}
\nc{\vf}{\vec f}
\nc{\vg}{\vec g}
\nc{\vh}{\vec h}
\nc{\vp}{\vec p}
\nc{\vq}{\vec q}
\nc{\vr}{\vec r}
\nc{\vs}{\vec s}
\nc{\vv}{\vec v}
\nc{\vw}{\vec w}
\nc{\vx}{\vec x}
\nc{\vy}{\vec y}
\nc{\vz}{\vec z}

\nc{\vB}{\vec B}
\nc{\vC}{\vec C}
\nc{\vD}{\vec D}
\nc{\vE}{\vec E}
\nc{\vF}{\vec F}
\nc{\vG}{\vec G}
\nc{\vH}{\vec H}
\nc{\vP}{\vec P}
\nc{\vQ}{\vec Q}
\nc{\vR}{\vec R}
\nc{\vS}{\vec S}
\nc{\vV}{\vec V}
\nc{\vW}{\vec W}
\nc{\vX}{\vec X}
\nc{\vY}{\vec Y}
\nc{\vZ}{\vec Z}

\nc{\ol}{\overline}
\nc{\abar}{\ol{a}}
\nc{\bbar}{\ol{b}}
\nc{\cbar}{\ol{c}}
\nc{\dbar}{\ol{d}}
\nc{\ebar}{\ol{e}}
\nc{\fbar}{\ol{f}}
\nc{\gbar}{\ol{g}}
\nc{\ibar}{\ol{\imath}}
\nc{\jbar}{\ol{\jmath}}
\nc{\kbar}{\ol{k}}
\nc{\lbar}{\ol{l}}
\nc{\mbar}{\ol{m}}
\nc{\nbar}{\ol{n}}
\nc{\pbar}{\ol{p}}
\nc{\qbar}{\ol{q}}
\nc{\rbar}{\ol{r}}
\nc{\sbar}{\ol{s}}
\nc{\ubar}{\ol{u}}
\nc{\vbar}{\ol{v}}
\nc{\wbar}{\ol{w}}
\nc{\xbar}{\ol{x}}
\nc{\ybar}{\ol{y}}
\nc{\zbar}{\ol{z}}

\nc{\Abar}{\ol{A}}
\nc{\Bbar}{\ol{B}}
\nc{\Cbar}{\ol{C}}
\nc{\Dbar}{\ol{D}}
\nc{\Ebar}{\ol{E}}
\nc{\Fbar}{\ol{F}}
\nc{\Jbar}{\ol{J}}
\nc{\Kbar}{\ol{K}}
\nc{\cKbar}{\ol{\cK}}
\nc{\Lbar}{\ol{L}}
\nc{\cLbar}{\ol{\cL}}
\nc{\Mbar}{\ol{M}}
\nc{\Nbar}{\ol{N}}
\nc{\Pbar}{\ol{P}}
\nc{\Qbar}{\ol{Q}}
\nc{\Rbar}{\ol{R}}
\nc{\Sbar}{\ol{S}}
\nc{\Tbar}{\ol{T}}
\nc{\Ubar}{\ol{U}}
\nc{\Vbar}{\ol{V}}
\nc{\cVbar}{\ol{\cV}}
\nc{\Wbar}{\ol{W}}
\nc{\cWbar}{\ol{\cW}}
\nc{\Xbar}{{\overline X}}
\nc{\Ybar}{{\overline Y}}
\nc{\Zbar}{{\overline Z}}
\nc{\cZbar}{{\overline \cZ}}

\nc{\epsbar}{\ol{\epsilon}}
\nc{\albar}{\ol{\al}}
\nc{\Albar}{\ol{\Al}}
\nc{\betabar}{\ol{\beta}}
\nc{\Betabar}{\ol{\Beta}}
\nc{\lambar}{\ol{\lambda}}
\nc{\kapbar}{\ol{\kappa}}
\nc{\zetabar}{\ol{\zeta}}
\nc{\Zetabar}{\ol{\Zeta}}
\nc{\taubar}{\ol{\tau}}
\nc{\Taubar}{\ol{\Tau}}
\nc{\psibar}{\ol{\psi}}
\nc{\Psibar}{\ol{\Psi}}
\nc{\tpsibar}{\ol{\tpsi}}
\nc{\tPsibar}{\ol{\tPsi}}
\nc{\phibar}{\ol{\phi}}
\nc{\Phibar}{\ol{\Phi}}
\nc{\chibar}{\ol{\chi}}
\nc{\mubar}{\ol{\mu}}
\nc{\nubar}{\ol{\nu}}
\nc{\rhobar}{\ol{\rho}}
\nc{\ombar}{\ol{\om}}
\nc{\Ombar}{\ol{\Om}}
\nc{\Deltabar}{\ol{\Delta}}
\nc{\Thetabar}{\ol{\Theta}}
\nc{\xibar}{\ol{\xi}}
\nc{\Xibar}{\ol{\Xi}}

\nc{\Dthbar}{\ol{\rm D3}}

\nc{\fdot}{\dot{f}}
\nc{\gdot}{\dot{g}}
\nc{\pdot}{\dot{p}}
\nc{\qdot}{\dot{q}}
\nc{\rdot}{\dot{r}}
\nc{\sdot}{\dot{s}}
\nc{\tdot}{\dot{t}}
\nc{\udot}{\dot{u}}
\nc{\vdot}{\dot{v}}
\nc{\wdot}{\dot{w}}
\nc{\xdot}{\dot{x}}
\nc{\xddot}{\ddot{x}}
\nc{\ydot}{\dot{y}}
\nc{\zdot}{\dot{z}}
\nc{\yddot}{\ddot{y}}

\nc{\Adot}{\dot{A}}
\nc{\Bdot}{\dot{B}}
\nc{\Cdot}{\dot{C}}
\nc{\Udot}{\dot{U}}
\nc{\Vdot}{\dot{V}}
\nc{\Wdot}{\dot{W}}

\nc{\taudot}{\dot{\tau}}
\nc{\phidot}{\dot{\phi}}
\nc{\psidot}{\dot{\psi}}
\nc{\chidot}{\dot{\chi}}
\nc{\sinp}{s_{\phi}}
\nc{\cosp}{c_{\phi}}
\nc{\tanp}{t_{\phi}}
\nc{\spone}{s_{\phi_1}}
\nc{\cpone}{c_{\phi_1}}
\nc{\tpone}{t_{\phi_1}}
\nc{\sptwo}{s_{\phi_2}}
\nc{\cptwo}{c_{\phi_2}}
\nc{\tptwo}{t_{\phi_2}}
\nc{\spth}{s_{\phi_3}}
\nc{\cpth}{c_{\phi_3}}
\nc{\tpth}{t_{\phi_3}}
\nc{\calp}{c_{\al}}
\nc{\salp}{s_{\al}}

\nc{\csch}{{\rm csch}}
\nc{\sech}{{\rm sech}}

\nc{\cothzlami}{\coth(z-\lam_i)}
\nc{\coshzlami}{\cosh(z-\lam_i)}
\nc{\sinhzlami}{\sinh(z-\lam_i)}

\nc{\cothzlamj}{\coth(z-\lam_j)}
\nc{\coshzlamj}{\cosh(z-\lam_j)}
\nc{\sinhzlamj}{\sinh(z-\lam_j)}

\nc{\cothlamij}{\coth(\lam_i-\lam_j)}
\nc{\coshlamij}{\cosh(\lam_i-\lam_j)}
\nc{\sinhlamij}{\sinh(\lam_i-\lam_j)}

\nc{\bah}{{\mathbf {\hat{A}}}}
\nc{\bX}{{\mathbf X}}
\nc{\ba}{{\bf a}}
\nc{\bb}{{\bf b}}
\nc{\bc}{{\bf c}}
\nc{\bd}{{\bf d}}
\nc{\bg}{{\bf g}}
\nc{\bk}{{\bf k}}
\nc{\bl}{{\bf l}}
\nc{\bm}{{\bf m}}
\nc{\bn}{{\bf n}}
\nc{\bo}{{\bf o}}
\nc{\bp}{{\bf p}}
\nc{\bq}{{\bf q}}
\nc{\br}{{\bf r}}
\nc{\bs}{{\bf s}}
\nc{\bt}{{\bf t}}
\nc{\bu}{{\bf u}}
\nc{\bv}{{\bf v}}
\nc{\bw}{{\bf w}}
\nc{\bx}{{\bf x}}
\nc{\by}{{\bf y}}
\nc{\bz}{{\bf z}}
\nc{\bom}{{\bf \om}}
\nc{\bombar}{{\mathbf \ombar}}
\nc{\bPhi}{{\bf \Phi}}

\nc{\rma}{{\rm a}}
\nc{\rmb}{{\rm b}}
\nc{\rmc}{{\rm c}}
\nc{\rmd}{{\rm d}}
\nc{\rmg}{{\rm g}}
\nc{\rk}{{\rm k}}
\nc{\rml}{{\rm l}}
\nc{\rmm}{{\rm m}}
\nc{\rmn}{{\rm n}}
\nc{\rmo}{{\rm o}}
\nc{\rmp}{{\rm p}}
\nc{\rmq}{{\rm q}}
\nc{\rmr}{{\rm r}}
\nc{\rms}{{\rm s}}
\nc{\rmt}{{\rm t}}
\nc{\rmu}{{\rm u}}
\nc{\rmv}{{\rm v}}
\nc{\rmw}{{\rm w}}
\nc{\rmx}{{\rm x}}
\nc{\rmy}{{\rm y}}
\nc{\rmz}{{\rm z}}

\nc{\dal}{\dot{\al}}
\nc{\thadot}{\dot{\tha}}
\nc{\thab}{\bar{\theta}}
\nc{\thal}{\theta^{\al}}
\nc{\thdal}{\bar{\theta}^{\dal}}

\nc{\thsigthm}{\tha \sigma^m \thab}
\nc{\thsigthn}{\tha \sigma^n \thab}

\nc{\Dal}{D_{\al}}
\nc{\Ddal}{\bar{D}_{\dal}}
\nc{\CDal}{{\cal D}_{\al}}
\nc{\CDdal}{\bar{\cal D}_{\dal}}

\nc{\eq}[1]{{(\ref{#1})}}
\nc{\eqtwo}[2]{{(\ref{#1},\ref{#2})}}
\nc{\eqthree}[3]{(\ref{#1},\ref{#2},\ref{#3})}
\nc{\eqfour}[4]{(\ref{#1},\ref{#2},\ref{#3},\ref{#4})}
\nc{\eqfive}[5]{(\ref{#1},\ref{#2},\ref{#3},\ref{#4,\ref{#5}})}
\nc{\non}{\nonumber}
\nc{\Fzero}{F_{(0)}}
\nc{\Ftwo}{F_{(2)}}
\nc{\Ffour}{F_{(4)}}
\nc{\Fone}{F_{(1)}}
\nc{\Fthree}{F_{(3)}}
\nc{\Ffive}{F_{(5)}}
\nc{\Fn}{F_{(n)}}
\nc{\Fp}{F_{(p)}}

\nc{\tFzero}{\tF_{(0)}}
\nc{\tFtwo}{\tF_{(2)}}
\nc{\tFfour}{\tF_{(4)}}
\nc{\tFone}{\tF_{(1)}}
\nc{\tFthree}{\tF_{(3)}}
\nc{\tFfive}{\tF_{(5)}}
\nc{\tFn}{\tF_{(n)}}
\nc{\tFp}{\tF_{(p)}}

\nc{\Czero}{C_{(0)}}
\nc{\Ctwo}{C_{(2)}}
\nc{\Cfour}{C_{(4)}}
\nc{\Cone}{C_{(1)}}
\nc{\Cthree}{C_{(3)}}
\nc{\Cfive}{C_{(5)}}
\nc{\Cn}{C_{(n)}}

\nc{\equ}{{\rm eq}}
\def\Im{{\rm Im \hspace{0.5mm} }}

\def\Re{{\rm Re \hspace{0.5mm}}}

\nc{\vol}{{\rm vol}}
\nc{\Ainf}{A_{\infty}}
\nc{\End}{{\rm End}}
\nc{\Ext}{{\rm Ext}}
\nc{\IIB}{{\rm IIB}}
\nc{\Ad}{{\rm Ad}}
\nc{\IIA}{{\rm IIA}}
\nc{\AdS}{{\rm AdS}}
\nc{\CFT}{{\rm CFT}}
\nc{\diag}{{\rm diag}}
\nc{\Log}{{\rm Log}}
\nc{\Dslash}{\ensuremath \raisebox{0.025cm}{\slash}\hspace{-0.32cm} D}
\nc{\cDslash}{\ensuremath \raisebox{0.025cm}{\slash}\hspace{-0.32cm} \cD}
\nc{\omslash}{\om\!\!\!/}
\nc{\no}{\!:\!\!}
\nc{\ointdz}{\oint\frac{dz}{2\pi i}}
\nc{\ointdzone}{\oint\frac{dz_1}{2\pi i}}
\nc{\ointdztwo}{\oint\frac{dz_2}{2\pi i}}
\nc{\ointdzb}{\oint\frac{d\zbar}{2\pi i}}
\nc{\ointdzbone}{\oint\frac{d\zbar_1}{2\pi i}}
\nc{\ointdzbtwo}{\oint\frac{d\zbar_2}{2\pi i}}
\nc{\dz}{\frac{dz}{2\pi i}}
\nc{\dzb}{\frac{d\zbar}{2\pi i}}
\nc{\bpm}{\begin{pmatrix}}
\nc{\epm}{\end{pmatrix}}
 \nc{\bitem}{\begin{itemize}}
 \nc{\eitem}{\end{itemize}}
 \nc{\exercise}{\vskip 2mm \noindent {\bf Exercise:}}
 \nc{\definition}{\vskip 2mm \noindent {\bf Definition:}}

\newcommand{\mathsym}[1]{{}}

\newcommand{\mc}{\mathcal}

\newcommand{\im}{\mathrm{Im}}
\newcommand{\re}{\mathrm{Re}}

\begin{document}

\begin{flushright}
SPIN-13/22 \\
ITP-UU-13/30
\end{flushright}

\vspace{0.5cm}
\begin{center}
\baselineskip=13pt {\LARGE \bf{Supersymmetric Black Holes in $AdS_4$\\ from Very Special Geometry \\}}
 \vskip1.5cm 
Alessandra Gnecchi$^{\dagger}$ and Nick Halmagyi$^{*}$ \\ 
\vskip0.5cm
$\dagger$\textit{ Institute for Theoretical Physics and Spinoza Institute, \\
Utrecht University, 3508 TD Utrecht, \\
The Netherlands} \\
\vskip 0.5cm
$^{*}$\textit{Laboratoire de Physique Th\'eorique et Hautes Energies,\\
Universit\'e Pierre et Marie Curie, CNRS UMR 7589, \\
F-75252 Paris Cedex 05, France}\\
\vskip0.5cm
A.Gnecchi@uu.nl \\
halmagyi@lpthe.jussieu.fr \\ 

\end{center}
\vskip1cm

\begin{abstract}
Supersymmetric black holes in AdS spacetime are inherently interesting for the AdS/CFT correspondence. Within a four dimensional gauged supergravity theory coupled to vector multiplets, the only analytic solutions for regular, supersymmetric, static black holes in AdS$_4$ are those in the STU-model due to Cacciatori and Klemm. We study a class of $U(1)$-gauged supergravity theories coupled to vector multiplets which have a cubic prepotential, the scalar manifold is then a {\it very special K\"ahler} manifold. When the resulting very special K\"ahler manifold is a homogeneous space, we find analytic solutions for static, supersymmetric AdS$_4$ black holes with vanishing axions. The horizon geometries of our solutions are constant curvature Riemann surfaces of arbitrary genus.
\end{abstract} 

\newpage

\section{Introduction}

Black holes in AdS space have been studied extensively since the development of the AdS/CFT correspondence \cite{Maldacena:1997re, Witten:1998qj, Gubser:1998bc} but supersymmetric black holes in AdS spacetime have proved to be rare finds. In four dimensional gauged supergravity coupled to $n_v$ vector multiplets, the only analytic solutions of regular, static, supersymmetric black holes are due to Cacciatori and Klemm (CK) \cite{Cacciatori:2009iz} and preserve two real supercharges. These solutions are found within the so-called STU supergravity model, which is standard nomenclature for a model with $n_v=3$. In this work we use the tools of special geometry and the general structure of the CK solution to find solutions within a particular infinite family of $\cN=2$ gauged supergravity theories.

The Lagrangian of four dimensional $\cN=2$ supergravity coupled to $n_v$-vector mulitplets is governed by special K\"ahler geometry \cite{deWit:1984pk, Strominger:1990pd, Castellani:1990zd}. When this geometry is in turn derived from a cubic prepotential
\be
F=-\frac{d_{ijk} X^i X^j X^k}{X^0}
\ee 
it is called {\it very special K\"ahler geometry} and this is focus of our work. Before gauging, such supergravity theories can be obtained by dimensional reduction from $\cN=2$ supergravity in five dimensions \cite{Gunaydin:1983bi}.

There is an additional simplification we will employ which facilitates the calculations, namely that $\cM_v$ be a homogeneous space. The central utility of this assumption is that it ensures the existence of a constant tensor $\hd^{ijk}$ (see appendix \ref{app:homo}. for its definition and numerous identities which it satisfies).
We will find that with this assumption the supersymmetric black hole equations are solvable in quite some generality. In fact the homogeneous, very special K\"ahler geometries have been classified some time ago in the nice work by de Wit and Van Proeyen \cite{deWit:1991nm,deWit:1992wf, deWit:1993rr} and includes several infinite families as well as certain sporadic geometries related to the dimensional reduction of the magical supergravity theories in five dimensions.

The R-symmetry of four dimensional $\cN=2$ supergravity is 
\be
 SU(2)_R\times U(1)_R
 \ee
and we are interested in gauging a $U(1)$ subgroup embedded as
\be
U(1)_g\subset SU(2)_R\ .
\ee 
This goes by the moniker FI-supergravity since the gauge couplings generate a potential much like Fayet-Iliopoulos terms in field theory \cite{Andrianopoli:1996vr}. A useful feature of this abelian gauging is that the scalar fields of the vector multiplets remain neutral under the gauged $U(1)_g$ vector\footnote{in addition, if hypermultiplets are present they remain decoupled}. The fermionic fields are minimally coupled and acquire a charge under the gauge field so that in addition to the abelian charges of the black hole $(p^{\Lambda}\,,q_{\Lambda}$), there are now additional parameters determining the theory, proportional to the gauging coupling $g$.

 To provide a duality covariant treatment we will consider the general case where the gauging is specified by a symplectic vector containing both electric and magnetic parameters
\begin{eqnarray}
\mc G^T=\left(g^{\Lambda},g_{\Lambda}\right)\ .
\end{eqnarray}
The supersymmetric Lagrangian for gauged $\mc N=2$ Supergravity has been constructed for electric gauging \cite{Andrianopoli:1996cm}, and it has been extended to magnetic gauging in the formalism of conformal supergravity \cite{deWit:2011gk}. In order to have a standard Lagrangian formulation that includes magnetic gauging, under which the fermions will be minimally coupled, one must also introduce auxiliary tensor fields.  Our strategy is to work with a symplectic completion of the BPS equations which results from electrically gauged models \cite{Dall'Agata:2010gj}%
\footnote{see also \cite{Hristov:2012nu} where a similiar formalism has been used}.

Early work on supersymmetric black holes in AdS space were suggestive of a no-go theorem prohibiting regular, half-BPS, asymptotically AdS$_4$ black holes \cite{Sabra:1999ux, Chamseddine:2000bk, Caldarelli1999}, for example  the black hole of \cite{Duff:1999gh} has a naked singularity. An early workaround was found in \cite{Caldarelli1999} where it was found that one could analytically continue AdS-Schwarzchild and construct a quarter-BPS solution of $\cN=2$ gauged supergravity with constant scalar fields with the proviso that the horizon is hyperbolic%
\footnote{which can then be quotiented by a discrete group to give a Riemann surface of genus $g>1$}.
Some time later, Cacciatori and Klemm \cite{Cacciatori:2009iz} successfullly demonstrated that by allowing for non-constant scalar fields, the solution of \cite{Caldarelli1999} admits a vast generalization within the $STU$-model  of $\mc N=2$  FI-gauged supergravity including solutions with spherical and flat hoirzons (see also \cite{Dall'Agata:2010gj,Hristov2011} for additional analysis of these BPS black holes%
\footnote{Further work has been done extending these solutions to non-BPS and non-extremal black holes \cite{Klemm:2012yg, Toldo:2012ec, Klemm:2012vm,Gnecchi:2012kb, Barisch:2011ui, Donos2012a}. There has also been recent work on supersymmetric AdS$_4$ black holes with hypermultiplets \cite{Donos:2012sy, Halmagyi:2013sla} where the resulting solutions are numerical.}).  
In a particular symplectic frame which will be elaborated on below, the CK solutions have four magnetic charges for the four gauge fields and the BPS Dirac quantization condition reduces this to three independant magnetic charges. The absence of electric charges is ultimately tantamount to the absence of axions in the CK solutions. The far-reaching work of Maldacena and Nunez \cite{Maldacena:2000mw} provides a framwork to understand the M-theory embedding of the CK solutions.

The central result of our current work is to derive analytic solutions for quarter-BPS black holes in AdS$_4$ which generalize the CK solution from the STU-model to models whose scalar manifold is a homogeneous very special K\"ahler manfold. Our new solutions also have vanishing axions and in the symplectic frame where the gaugings are electric, the charges are all magnetic.  A first step towards this result was the work \cite{Halmagyi:2013qoa} where these models were studied and the general solution for supersymmetric horizon geometries of the form $AdS_2\times \Sig_g$ was found. That solution allows for generic gaugings and both electric and magnetic chargs whereas the black hole solutions of the current work will be far more restrictive. Regardless, the results of \cite{Halmagyi:2013qoa} constitute a solution to the IR boundary conditions for our black holes. In the current work we also analyze the UV AdS$_4$ boundary conditions and find that they are equivalent to the supersymmetric attractor equations in {\it ungauged} supergravity, before proceeding to solve for the entire black hole. A key step in our argument is to show that for the static black hole ansatz, a solution with vanishing axions puts strong constraints on the allowed gauging parameters.

Our paper is organised as follows. In section \ref{secgeneralitites}. we review some basic facts about $\cN=2$ gauged supergravity in the formalism of \cite{Dall'Agata:2010gj}, the black hole ansatz and the resulting BPS equations. In section \ref{sec:AdS4UV}. we solve the UV boundary conditions; we give the explicit solution for AdS$_4$ solutions in $\cN=2$ FI-gauged supergravity. In section \ref{sec:BHspecial}. we perform our central calculation; an analytic solution for axion-free black holes in models whose scalar manifold is a homogeneous very special K\"ahler geometry. Section \ref{sec:AdS2IR}. contains  some comments about the IR boundary conditions and regularity of the solutions.

\section{Generalities of $\frac14$-BPS static black holes in $AdS_4$}\label{secgeneralitites}

The Lagrangian of $\cN=2$ gauged supergravity coupled to $n_v$ vector multiplets is  
 \begin{eqnarray}\label{N2action}
S_{4d}=\int\,d^4x\sqrt{-g} \left( \frac{1}{2} R- g_{ij} \p_{\mu} z^i\p^{\mu} z^j + \mc I_{\Lambda\Sigma}F^{\Lambda}_{\mu \nu}F^{\Sigma\,\mu \nu}+ \cR_{\Lam \Sig} \eps^{\mu\nu\rho\sig} F_{\mu\nu}^\Lam F_{\rho\sig}^\Sig-V_g
\right)\ .
\end{eqnarray}
We will work in the symplectically covariant formulation of \cite{Dall'Agata:2010gj} which for the black hole ansatz we use, provides a covariant form of the BPS equations. Our spacetime ansatz is that of a static black hole with constant curvature horizon:
\be
ds_4^2= -e^{2U}dt^2 + e^{-2U} dr^2 + e^{2(V-U)} d\Sig_g^2\,,
\ee
where $d\Sig_g$ is the uniform metric on $\Sig_g=\{S^2,T^2,\HH^2/\Gamma\}$ of curvature $\kappa=\{1,0,-1\}$ respectively\footnote{The discrete group $\Gamma$ is a Fuchsian group and its precise form does not alter this local analysis}. The gauge fields are chosen such that 
\be
p^\Lam = \frac{1}{\vol(\Sig_g)}\int_{\Sig_g} F^\Lam\,,\quad\quad q_\Lam = \frac{1}{\vol(\Sig_g) }\int_{\Sig_g} G_\Lam
\ee
where
\be
G_\Lam=\frac{\delta L_{4d}}{F^\Lam} = \cR_{\Lam \Sig} F^\Sig + \cI_{\Lam\Sig} * F^\Sig
\ee
is the dual field strength and $\vol(\Sig_g)$ is the volume of $\Sig_g$. In fact the BPS equations are independant of the precise profiles for the gauge fields, they depend only on the charges $(p^\Lam, q_\Lam)$. The scalar fields depend only on the radial co-ordinate $z^i=z^i(r)$.

As mentioned in the introduction the gauging is parametrized by a symplectic vector $\cG$, corresponding to the gravitino charges under the $U(1)$ field of the gauging. So our data is organised into a pair of symplectic vectors: 
\be
\cQ= \bpm p^\Lam \\ q_\Lam \epm\,,\quad\quad\cG= \bpm g^\Lam \\ g_\Lam \epm\,.
\ee
In our notation, the symplectic section over the scalar manifold $\cM_v$ is denoted\footnote{We use the notation and conventions of \cite{Andrianopoli:1996vr} as much as possible, apart from the signature of space-time which we take to be {\it mostly plus}.} $\cV$:
\be
\cV=\bpm L^\Lam \\ M_\Lam\epm = e^{K/2}\bpm X^\Lam \\ F_\Lam \epm
\ee
and we have used the symplectic inner product between two  vectors $A=(A^{\Lambda},A_{\Lambda})$ and $B=(B^{\Lambda},B_{\Lambda})$
\be
\langle A ,B \rangle \equiv A^T\Omega B = B^\Lam A_\Lam - A^\Lam B_\Lam 
\ee
to produce the invariants
\be
\cZ=\langle  \cQ,\cV \rangle\,,\ \ \ \ \ \cL=\langle  \cG,\cV \rangle\,,\ \ \ \ \ 
\cZ_i=\langle  \cQ,D_i\cV \rangle\,,\ \ \ \ \ \cL_i=\langle  \cG,D_i\cV \rangle\,.
\ee

The BPS equations for preservation of at least two supercharges were derived in \cite{Cacciatori:2009iz} for electric gaugings and \cite{Dall'Agata:2010gj} for general dyonic gaugings. We use the results of \cite{Dall'Agata:2010gj} which were found by reducing \eq{N2action} to one dimension and re-writing the resulting action as a sum of squares. The final result gives the BPS equations to be:
\bea
2e^{2V} \del_r \Bslb \Im \blp e^{-i\psi} e^{-U} \cV\brp \Bsrb &=&8e^{2(V-U)} \Re(e^{-i\psi} \cL) \Re(e^{-i\psi} \cV) - \cQ -e^{2(V-U)} \Om \cM \cG \label{delcV}\\
\del_r (e^V) &=& 2e^{V-U} \Im ( e^{-i\psi} \cL ) \label{Vdot} \\
\psi'&=& -A_r -2 e^{-U} \Re(e^{-i\psi} \cL) \label{psidot}
\eea
The connection $A_\mu$ is given by
\be
A_\mu=\Im\blp  \del_\mu z^i \del_i K\brp
\ee
and the matrix $\cM$ is given in \eq{Mdef}. When $\cM$ is contracted with the symplectic form $\Om$ it gives a complex structure on the $Sp(2n_v+2,\RR)$ bundle over $\cM_v$:
\be
\Om \cM \cV = -i\cV \,,\quad\quad \Om \cM ( D_i\cV )= iD_i\cV\,.
\ee
While \eq{delcV} may seem cumbersome, it is just a repackaging of the first order equations for the scalar fields $z^i$ and the metric function $e^U$. This repackaging is useful since much like the ungauged $\cN=2$ supersymmetric black holes \cite{Behrndt:1997ny, Denef:2000nb}, the analytic black hole solutions are particularly simple when expressed in terms of this data. By re-deriving a version of these equations in a frame with electric gaugings using the formulae of \cite{Andrianopoli:1996vr} one can establish that the resulting black holes preserve two out of eight real supercharges along the flow and four at the horizon. 

Notice that \eqref{psidot} is the equation for the phase $\psi$ of the supersymmetry parameter. This is not a new degree of freedom and in fact one can show \cite{Dall'Agata:2010gj} that this is given algebraically by the phase of a superpotential $W=e^Ue^{-i\psi}(\cZ-i e^{2(V-U)}\cL)$, or equivalently
\be
e^{2i\psi}= \frac{\cZ-i e^{2(V-U)}\cL}{\cZbar+ i e^{2(V-U)}\cLbar}\,.
\ee
Using this definition the flow equation \eqref{psidot} follows from \eqref{delcV} and \eqref{Vdot}.
Since the gravitino is charged, there is a Dirac quantization condition 
\be
\langle \cG,\cQ\rangle \in \ZZ
\ee
and the supersymmetry conditions fix this integer to be the curvature of the horizon:
\be\label{BPSDirac}
\langle \cG,\cQ\rangle =-\kappa\,.
\ee
It is interesting to note that \eq{BPSDirac} is the only place where the curvature of the horizon geometry appears. Pragmatically this means that solutions are independant of the curvature of $\Sig_g$ but the regularity conditions do
depend on $\kappa$. 

The single center, static black holes we consider in this work interpolate between $AdS_4$ at large $r$ and $AdS_2\times \Sig_g$ at some finite positive $r=r_h$. The metric functions for these spaces is 
\bea
AdS_4:&& e^{U}=\frac{r}{R}\,,\ \ \ \ e^V=\frac{r^2}{R} \label{AdS4ansatz}\\
AdS_2\times \Sig_g:&& e^{U}=\frac{r}{R_2}\,,\ \ \ \ e^V=\frac{rR_2}{R_1} \label{AdS2SigAnsatz}
\eea
and the scalar fields and the phase $\psi$ are constant. In the next section we analyze the $AdS_4$ solutions as a function of the gauging parameters. The spectrum of horizon geometries \eq{AdS2SigAnsatz} as a function of both the gaugings $(g^\Lam,g_\Lam)$ and charges $(p^\Lam,q_\Lam)$ was solved in \cite{Halmagyi:2013qoa}.

\section{UV boundary conditions from Very Special Geometry} \label{sec:AdS4UV}

In this section we solve the BPS equations \eq{delcV} and \eq{Vdot} for AdS$_4$ geometries \eq{AdS4ansatz}, constant scalars and vanishing charges. We do not impose \eq{BPSDirac}. This allows us to identify the subspace of gauging parameters which is needed for black holes with vanishing axions.

\subsection{General $AdS_4$ solutions}
We first analyze the boundary conditions in the UV, where we can obtain the exact solution to the BPS equations.
For $AdS_4$, the metric functions are given by \eq{AdS4ansatz}, the scalars and the phase $\psi$ are constant and the charges are zero:
\be
z=x_0+i y_0\,,\ \ \ \psi=\psi_0\,,\ \ \ \ \cQ=0\,.
\ee
With this ansatz, the equations give
\bea
\cG&=&-2\Im \Bslb \cLbar \cV \Bsrb  \label{GVEq1}\\
\cL&=&\re\cL + i \im\cL= \frac{i}{R} e^{i\psi_0} \label{GVEq2}
\eea
These equations are in fact identical to the attractor equations for solutions of the form $AdS_2\times S^2$ in {\it ungauged} $\N=2$ supergravity \cite{Ferrara:1996dd} with the obvious replacement of the gauging parameters $\cG$ with charges $\cQ$. 

When $\cM_v$ is a very special geometry, these equations are quite tractable and have been analyzed in \cite{Shmakova:1996nz}. In special co-ordinates \eq{GVEq1} ammounts to
\bea
g^0 &=& 2 e^{K/2} \,\im\cL \,, \label{AdS4Eq1}\\
g_0&=& g^0 d_{ijk} (x^i x^j x^k -3 x^i y^j y^k) + 2 \, \re\cL\, e^{K/2} d_{ijk} (y^i y^j y^k - 3 y^i x^j x^k) \,, \\
g^i&=& g^0 x^i - 2 e^{K/2}\, \re\cL \, y^i \,,\\
g_i&=& 3 g^0 d_{ijk} (y^j y^k - x^j x^k)+ 12\, \re\cL\, e^{K/2} d_{ijk} x^j y^k\label{AdS4Eq4}
\eea
and the solution requires inverting these and expressing the scalars $(x^i,y^i)$ and $(\re\cL,\im\cL)$ in terms of the gaugings $(g^\Lam,g_\Lam)$. 

If one makes the assumption that $g^0=0$ the general solution is quite straightforward to obtain:
\bea
\frac{1}{R^2}&=&\sqrt{- 4 d_g g_0 + \frac{1}{3} (d_g^{-1})^{ij} g_i g_j} \label{g0Sol0} \\
x^i&=&- \frac{1}{6} (d_g^{-1})^{ij} g_j\,, \label{g0Sol1}\\
y^i&=& \frac{g^i}{2d_g} \frac{1}{R^2}\,, \label{g0Sol2} \\
\re\cL&=&  \frac{\eps}{R}\,, \label{g0Sol3} \\
\im\cL&=& 0\,,  \label{g0Sol4}\\
\psi_0&=&-\eps \frac{\pi}{2}\,.\label{g0Sol5}
\eea
where $\eps=\pm1$ is a convention. Up to obtaining an expression for $d_g^{-1}$, this comprises an explicit solution.

With $g^0\neq 0$ the general solution requires solving the set of $n_v$ real, quadratic equations
\be
\Delta_i= d_{ijk} \ty^j \ty^k \label{Deltadyy}
\ee
where
\bea
\Delta_i &=& 3 d_{ijk} g^j g^k + g^0 g_i\,,\ \ \ \ \ \ \ \ty^i=\sqrt{12} |\cL|e^{K/2}y^i \,. \label{Deltadef}
\eea
The general solution to \eq{Deltadyy} is not known and being real equations, they are not in general  guaranteed to have real solutions. If we assume $\cM_v$ to be a homogeneous space in addition to a very special K\"ahler geometry, then we can solve \eq{Deltadyy} using the {\it constant} tensor
\be
\hd^{ijk} =\frac{g^{il}g^{jm} g^{kn} d_{ijk}}{d_y^2}
\ee
which satisfies numerous identities detailed in appendix \ref{app:homo}. The solution to  \eq{g0Sol0}-\eq{g0Sol5} is then given by
\bea
\frac{1}{R^2}&=& \sqrt{\cI_4(g^\Lam,g_\Lam)} \label{gnotzero1}\\
\re\cL&=& \frac{2 \cI_2(g^\Lam,g_\Lam) \cI_4(g^\Lam,g_\Lam)^{1/4}}{\sqrt{ \cI_4(g^\Lam,g_\Lam)+4 \cI_{2}(g^\Lam,g_\Lam)^2}} \label{gnotzero2} \\
\im\cL&=&\frac{\bslb \cI_4(g^\Lam,q_\Lam)\bsrb^{3/4}}{ \sqrt{\cI_4(g^\Lam,g_\Lam)+4 \cI_{2}(g^\Lam,g_\Lam)^2}}\label{gnotzero3} \\
y^i&=& \frac{3}{32} \frac{1}{(g^0)^2} \frac{\cI_4(g^\Lam,g_\Lam)^{1/2}}{ \cI_4(g^\Lam,g_\Lam)+4 \cI_{2}(g^\Lam,g_\Lam)^2} \hd^{ijk} \Delta_j \Delta_k \label{gnotzero4} \\
x^i &=& \frac{g^i}{g^0} + \frac{3}{16} \frac{\cI_2(g^\Lam,g_\Lam)}{(g^0)^2} \frac{ \hd^{ijk} \Delta_j \Delta_k}{ \cI_4(g^\Lam,g_\Lam)+4 \cI_{2}(g^\Lam,g_\Lam)^2}\label{gnotzero5}
\eea
where we have used the identity
 \bea
\hd_\Delta&=& 16 (g^0)^2\bslb  \cI_4(g^\Lam,g_\Lam)+4 \cI_{2}(g^\Lam,g_\Lam)^2\bsrb
 \eea
 and the invariants $(\cI_2,\cI_4)$ are defined in \eq{quadratic} and \eq{quartic}. Note that even though we derived \eq{gnotzero1}-\eq{gnotzero5} assuming $g^0\neq0$, they have a smooth $g^0\ra 0$ limit which agrees with \eq{g0Sol0}-\eq{g0Sol5}.

\subsection{$AdS_4$ Solutions with vanishing axions}

The black holes we will study below all have vanishing axions and so we would first like to understand the space of $AdS_4$ solutions with vanishing axions. These will serve as the asymptotic UV boundary conditions for our black holes. There are $n_v$ constraints $x^i=0$ and from  \eq{AdS4Eq1}-\eq{AdS4Eq4} one finds that they take the form
\be\label{AdS4axionconstraint}
d_g g_i =-3 g_0 g^0\, d_{g,i}\,.
\ee
So in general we expect the space of zero-axion AdS$_4$ solutions to be $n_v+2$ dimensional. Assuming $g_0\neq 0$ the explicit solution is given by
\bea
y^i&=& -\sqrt{-\frac{g_0}{d_g}} \, g^i\,,\quad\quad \Re \cL= \sqrt{2} (-g_0 d_g)^{1/4}\,,\quad\quad \Im \cL= \frac{\sqrt{2} g_0 g^0 }{(-g_ 0 d_g)^{1/4}}\,.
\eea
The co-dimension one subspace with $(g^0,g_i)=(0,0)$ will be the focus of our work in the next section.

\section{Black holes from Very Special Geometry}\label{sec:BHspecial}

In this section we solve for supersymmetric black holes with vanishing axions. We restrict to black holes with $g^0=g_i=0$. We first demonstrate that for this class of black holes the phase $\psi$ is constant. We then proceed to make an ansatz for the rescaled section and solve analytically. To describe our ansatz we first fix the K\"ahler gauge by choosing special co-ordinates
\be
X^\Lam = \bpm 1 \\ z^i \epm \,,
\ee
where $z^i=x^i + i y^i$. Having done this, we assume that the axions vanish 
\be
x^i=0\,.
\ee
Note that since we have fixed that K\"ahler gauge, we cannot shift $\psi$.

\subsection{Constant $\psi$}
We now explore which configurations of gauge couplings result in a constant supersymmetry phase when the axions are set to zero. Combining \eq{GVEq2} woth \eq{psidot} we see that
\begin{eqnarray}
\psi'|_{\infty}&=&0\ .
\end{eqnarray}
We then proceed by induction in order of derivatives on $\psi$. 

The condition $x^i=0$ implies that $\mc A_r$ is zero along the whole flow but this is not enough to show that $\psi$ is constant, we will need assume one of the two configurations%
\footnote{In both these configurations one can use a duality transformation to set the remaining gauge couplings equal in magnitude, for example in the STU-model to the frame with $g_0=-g^i=g$ which has an M-theory lift. We find it simpler to refrain from making this transformation as it allows us to more easily maintain covariant formulae.}
\bea
1.&&g^0=g_i=0\,,\label{gauge1} \\
2.&& g_0=g^i=0\,.\label{gauge2}
\eea
For the case  \eq{gauge1} then \eq{psidot} reduces to
\begin{eqnarray}
\psi'&=&  -2 e^{-U(r)}\mc L(r)\cos(\psi(r))\ .
\end{eqnarray}
Since the differential equation is of the form 
\begin{eqnarray}
\psi(r)'&=&  a(r) \cos\psi(r) \nonumber\\
\psi(r\rightarrow\infty)&=&0\ ,
\end{eqnarray}
every $n$-th derivative of $\psi$ depends only on terms which are
\begin{itemize}
\item terms proportional to $\cos\psi(r)$, which vanishes at infinity since $\psi_{\infty}=\pm\pi/2$,
\item terms containing derivatives of $\psi(r)$ up to the order $n-1$, which vanish at infinity by asuumption
\end{itemize}
This means that all derivatives calculated at infinity are zero, and thus the phase $\psi$ is constant throughout the entire spacetime. The latter case of \eq{gauge2} goes through similarly but with $\Re(\cL)=0$ throughout. 

This does not exhaust the possible black holes with vanishing axions in these models since the UV asymptotics given by \eq{gauge1} and \eq{gauge2} are co-dimension one in the space defined by \eq{AdS4axionconstraint}. In simple examples we have found that in the UV we can use a duality transformation to generate a general AdS$_4$ solution satisfying \eq{AdS4axionconstraint} from one satisfying \eq{AdS4axionconstraint} and \eq{gauge1} but such a transformation does generate axions in the bulk of the flow. It would be interesting to solidify these observations and determine unambiguously whether or not vanishing axions implies a constant spinor for this entire class of black holes.

\subsection{\label{Sec-ansatz}The ansatz}

With a constant phase $\psi$, the BPS equations \eq{delcV} and \eq{Vdot} simplify somewhat:
\bea
2e^{2V} \del_r \Bslb \Im \blp e^{-i\psi_0} e^{-U}L^\Lam \brp \Bsrb &=& - p^\Lam + e^{2(V-U)} \cI^{\Lam \Sig} g_\Sig \label{ImLBPS1}\\
2e^{2V} \del_r \Bslb \Im \blp e^{-i\psi_0} e^{-U} M_\Lam \brp \Bsrb &=& - q_\Lam -e^{2(V-U)}  \cI_{\Lam \Sig} g^\Sig\label{ImLBPS2} \\
\del_r (e^V) &=& 2e^{V-U} \blp g_0 L^0 - g^i M_i\brp  \label{VEq}\,.
\eea
Despite the fact that in the UV we could solve the full $g^0=0$ solution space in all generality, to proceed further with the black hole solution we make the simplifying assumption that $\cM_v$ is a homogeneous space. So we assume that \eq{gauge1} holds and then continue by solving for $(L^0,M_i)$, we find the equations (\ref{ImLBPS1}-\ref{VEq}) become
\bea 
2  e^V \del_r\blp  \tL^0\brp-2 \del_r(e^V) \tL^0  &=& - p^0 - 8 g_0 (\tL^0)^2 \label{BPS1}\\
2e^{V}  \del_r\blp \tM_i\brp-2 \del_r(e^V) \tM_i  &=& - q_i - \Bslb  \frac{9}{4} d_{ijk} \hd^{klm} \tM_l \tM_m - 8 \tM_i \tM_j  \Bsrb g^j\label{BPS2} \\
\del_r (e^V) &=& 2\Bslb g_0 \tL^0 - g^i \tM_i \Bsrb\,,\label{VEq2}
\eea
where we have defined rescaled the sections
\bea
\tL^\Lam&=&e^{V-U} L^\Lam \,,\ \ \ \ 
\tM_\Lam=e^{V-U}M_\Lam\,
\eea
and have used the following data (which is true for $x^i=0$)
\bea
&& L^\Lam=e^{K/2} \bpm1 \\ iy ^i\epm\,,\ \ \ \ M_\Lam=e^{K/2} \bpm -i d_y \\ 3 d_{y,i} \epm\,,\\
&& \Om \cM=\bpm 0& -\cI^{-1}\\ \cI & 0\epm\,,\ \ \ \ \  \cI_{\Lam \Sig} =- d_y\bpm 1 & 0 \\ 0 & 4 g_{ij} \epm\,, \\
&& d_y g_{ij}= -\frac{9}{16}d_{ijk}\hd^{klm} M_l M_m +2 M_i M_j \,.
\eea
Due to our assumption that $g^0=0$, we see that $\Re(e^{-i\psi} \cL)=0$ and from \eq{psidot} we see that $\psi=\psi_0=-\pi/2$ is constant throughout the flow. Another consequnce of constant $\psi$ and $x^i=0$ is that from \eq{ImLBPS1} and \eq{ImLBPS2} one can show that $p_i=q^0=0$.

Once we have solved for $(e^V,\tL^0,\tM_i,\psi_0)$ we will obtain the scalar fields and $e^U$ using the identities
\bea
y^i &=& \frac{3}{8} \hd^{ijk} M_j M_k \,, \\
1&=& L^0 \hd^{ijk} M_i M_j M_k
\eea
which gives
\bea
y^i &=& \frac{3}{8}\frac{ \hd^{ijk} \tM_j \tM_k }{\sqrt{\tL^0 \hd^{lmp} \tM_l \tM_m \tM_p}}\,.  \label{yiId} \\
e^{4U}&=&\frac{e^{4V}}{\tL^0 \hd^{ijk} \tM_i \tM_j \tM_k}\,, \label{e4UId}
\eea

\subsection{The solution}

Taking the solution of \cite{Cacciatori:2009iz} as inspiration, we make the ansatz
\bea
e^V&=& \frac{r^2}{R} -v_0\,, \label{v0ansatz}\\
\tL^0&=& \al^0 r + \beta^0 \,, \label{L0ansatz}\\
\tM_i&=& \al_i r + \beta_i\,.\label{Miansatz}
\eea
This is a rather enlightened ansatz which is difficult to motivate in advance. In principle the UV boundary conditions fix $(R,\al^0,\al_i)$ and the IR boundary conditions in principle fix $(v_0,\beta^0,\beta_i)$. The flow equations will then highly overconstrain the system and in this sense it will be quite miraculous should the BPS equations admit solutions of such a simple form.

From \eq{VEq2} we get
\bea
\frac{2}{R}&=& 2 \al^0 g_0 - 2 \al_i g^i \label{alphasum}\\
0&=& g_0 \beta^0 - g^i \beta_i \label{betasum}
\eea
We find from \eq{BPS1}
\bea
\frac{2\al^0}{R} -\frac{4\al^0}{R}&=& -8g_0 (\al^0)^2 \\
-\frac{4}{R}\beta^0 &=& -16 g_0 \al^0 \beta^0 \\
-2 v_0 \al^0&=& - p^0 -8g_0 (\beta^0)^2
\eea
and we immediately see that
\be\fbox{$
\al^0= \frac{1}{4g_0R} 
$}\label{alpha0Sol}\ee
\be
\fbox{$\beta^0=\frac{\eps_0}{g_0}\sqrt{\frac{1}{8}\Blp\frac{v_0}{2R} - g_0 p^0\Brp}
$} \label{beta0sol} 
\ee
where $\eps_0=\pm1$.

 Then from \eq{BPS2} we get
\bea
-\frac{2\al_i}{R}&=& - \Bslb  \frac{9}{4} d_{ijk} \hd^{klm} g^j \al_l \al_m - 8 \al_i \al_j g^j \Bsrb \label{BPS3}\\
-\frac{4}{R} \beta_i&=& -  \Bslb  \frac{9}{2} d_{ijk} \hd^{klm} g^j \al_l \beta_m - 8 (\al_i \beta_j+ \beta_i \al_j ) g^j\Bsrb  \label{BPS4} \\
-2v_0 \al_i &=& -q_i - \Bslb  \frac{9}{4} d_{ijk} \hd^{klm} g^j \beta_l \beta_m - 8 \beta_i \beta_j g^j \Bsrb  \label{BPS5}
\eea
and from \eq{BPS3} we find that 
\be\fbox{$
\al_i =- \frac{3}{4 R d_g}d_{ijk} g^j g^k $}\ \ \ \Rightarrow \ \ \ \al_i g^i =- \frac{3}{4R}\,.
\label{alphaiSol}\ee
which agrees with the UV analysis in section \ref{sec:AdS4UV}. We can immediately see that \eq{alphasum} is satisfied and with some effort (using identities in appendix \ref{app:homo}) one can also compute that \eq{BPS4} is automatically satisfied.

It now remains to use \eq{betasum} and \eq{BPS5} to solve for $(v_0,\beta_i)$. This is $(n_v+1)$--equations for $(n_v+1)$--parameters and should thus admit a solution. From \eq{BPS5} we should get an expression for $\beta_i$:
\bea
-\frac{4}{9}(d_g^{-1})^{ij}q_j- \frac{2v_0}{3 R d_g}g^i&=&   \hd^{ilm}  \beta_l \beta_m - \frac{32}{9} (d_g^{-1})^{ij}\beta_j \beta_k g^k  \label{pi1}
\eea
Now using $\hd^{ijk}$ we have an explicit expression for $d_g^{-1}$
\bea
(d_g^{-1})^{ij}=\frac{1}{d_g}\Bslb \frac{27}{16} \hd^{ijk} d_{g,k}-3 g^i g^j \Bsrb
\eea
and we get that \eq{pi1} becomes
\bea
-\frac{4}{9}(d_g^{-1})^{ij}q_j- \frac{2v_0}{3 R d_g}g^i
&=& \hd^{ilm}  \Bslb \beta_l - \frac{3}{d_g} d_{g,l}  \beta_j g^j\Bsrb
  \Bslb \beta_m - \frac{3}{d_g} d_{g,m}  \beta_j g^j \Bsrb -\frac{32}{3d_g} g^i (\beta_ig^i)^2 \label{pi2}\,.
\eea
Now using \eq{betasum} and \eq{beta0sol} we know that
\be
\beta_i g^i=\eps_0\sqrt{\frac{1}{8}\Blp \frac{v_0}{2R} -g_0p^0\Brp }\label{betaigi}
\ee
and thus we can define an object $\Pi^i$ which depends only on $\cG$ and $\cQ$: 
\bea
\Pi^i&=&- \frac{3}{4d_g} \hd^{ijk} d_{g,k} q_j+ g^i \frac{4}{3d_g} (q_m g^m - g_0 p^0)
\eea
so that \eq{pi2} becomes a familiar type of equation (see eq. \eq{hdEq})
\be
\Pi^i= \hd^{ilm} \Bslb \beta_l - \frac{3}{d_g} d_{g,l}  \beta_i g^i\Bsrb
  \Bslb \beta_m - \frac{3}{d_g} d_{g,m} \beta_i g^i\Bsrb\,. 
\ee

This can be solved explicitly and we get an expression for $\beta_i$ in terms of $v_0$ along with the charges and gaugings:
\bea
 \beta_i =\eps \sqrt{\frac{27}{64}}\frac{d_{ijk} \Pi^j \Pi^k}{\sqrt{d_{\Pi}}}+ \frac{3\beta_j g^j}{d_g} d_{g,i}  \,.\label{betasol1}
\eea
where again $\eps_i=\pm1$. It remains to solve for $v_0$, which is done as follows. Contracting \eq{betasol1} with $g^i$ gives
\bea
\beta_i g^i &=&-\frac{\eps}{2}\sqrt{\frac{ 27}{64}}\frac{d_{ijk} g^i \Pi^j \Pi^k}{ \sqrt{d_\Pi}}
\eea
which when combined with \eq{betaigi} gives the solution for $v_0$ purely in terms of the charges and gaugings:
\be\fbox{$
v_0=2R\Bslb g_0 p^0 +\frac{ 27( d_{ijk} g^i \Pi^j \Pi^k)^2}{32 d_\Pi} \Bsrb
$}\label{v0sol}\ee
so that
\be\fbox{$
\beta_i= \eps\sqrt{\frac{27}{64 \, d_\Pi}} \Bslb  d_{ijk}  \Pi^j \Pi^k-\frac{3}{2 d_g}d_{g,i}  d_{lmn} g^l \Pi^m \Pi^n  \Bsrb 
$}\label{betaiSol}\ee
and
\be
\eps=-\eps_0\,.
\ee

\subsection{Constant scalar flows}

We conclude this analysis by writing out the universal black hole with constant scalar fields. This is well known to require a hyperbolic horizon \cite{Dall'Agata:2010gj, Caldarelli1999} and we confirm that result here. These flows with constant scalars have 
\bea
\beta^0&=&0\ \ \  \Rightarrow\ \ \ v_0=2R g_0 p^0 \label{v0SolConst}\\
\beta_i&=&0 \ \ \ \Rightarrow \ \ \ \Pi^i=0
\eea
The constraint $\Pi^i=0$ gives
\bea
q_i&=& -\frac{3 g_0 p^0}{d_g} d_{g,i} 
 \eea
Contracting with $g^i$ and using \eq{BPSDirac} gives
\bea
g^i q_i=-3 g_0 p^0\ \ \  \Rightarrow \ \ \  \kappa=-4 g_0 p^0
\eea
which must be positive by \eq{v0SolConst} and therefore these solutions require 
\be
\kappa=-1 \ \ \ \Rightarrow\ \ \ \Sig_g=\HH^2/\Gamma \,.
\ee

\subsection{Summary of the solution}
It may provide some clarity to provide the entire solution in one place.
The rescaled sections $(\tL^0,\tM_i)$ are given by \eq{L0ansatz} and \eq{Miansatz} in terms of $(\al^0,\beta_0,\al_i,\beta_i)$ which in turn are given in \eq{alpha0Sol},\eq{beta0sol},\eq{alphaiSol},\eq{betaiSol}. The metric function $e^V$ is given by \eq{v0ansatz} and \eq{v0sol}. To obtain the scalar fields $y^i$ and the metric function $e^U$ one uses \eq{e4UId} and \eq{yiId}. 

If one chooses $n_v=3$, $d_{123}=\frac{1}{6}$ and $\hd^{123}=\frac{32}{3}$ (and symmetric permutations) one obtains the so-called STU-model, the model employed in \cite{Cacciatori:2009iz}. For this model with in addition
\be
g_0=-g^i=g
\ee 
the AdS$_4$ black holes can be embedded into M-theory compactified on $S^7$ \cite{deWit:1981eq, Duff:1999gh, Cvetic1999b}.  For more general  very special K\"ahler manifolds which are homogeneous spaces, one can find the explicit form of the corresponding $d_{ijk}$-tensor in section 5. of \cite{deWit:1992wf}. The embedding of these models into M-theory or string theory remains an important oustanding problem.

\subsection{Rotation to electric gaugings}
If the reader is for whatever reason uncomfortable with the use of magnetic gauging parameters, one can rotate the solutions of this paper to a frame where the gaugings are electric. Explicitly one finds that from the prepotential \eq{Prepotential} and the gauging parameters $(g_0,g^i)$ one can rotate to a new symplectic frame using 
\be
\cS=\bpm A& B \\ C & D\epm\,,\ \ \ A=D=\diag\{1,0,\ldots ,0\}\,,\ \ \ B=-C=\diag\{0,1\ldots ,1\}
\ee
to find a new prepotential
\bea
\tF=-i\sqrt{\frac{1}{16}} \sqrt{ \tX^0\hd^{ijk} (\delta_{il}\tX^l)(\delta_{jm}\tX^m)(\delta_{kn}\tX^n)}
\eea
with new gaugings
\be
\tg_i= - g^i \,,\ \ \ \tg_0=g_0\,.
\ee
The space-time metric and scalar fields remain invariant under this symplectic transformation. For the STU model this frame has $\tF=-2i\sqrt{\tX^0\tX^1 \tX^2 \tX^3}$ and this is the frame which is used to embed the STU model into the de Wit-Nicolai theory \cite{deWit:1982ig} and thus into M-theory.

\section{$AdS_2\times \Sig_g$: IR boundary conditions}\label{sec:AdS2IR}
We now make some brief statements about regularity of our solutions. To map out the subspace of regular solutions from section \ref{sec:BHspecial}. one needs to ensure that the scalar fields do not vanish before the horizon is reached:
\be
r_h>r_i\,,\quad\quad r_h> r_0
\ee
where 
\be
r_h=\sqrt{v_0R}\,,\quad\quad r_0 = -\frac{\beta^0}{\al^0}\,,\quad\quad r_i=-\frac{\beta^i}{\al^i}\,.
\ee
From $r_h>r_0$ we get
\be
\sqrt{v_0R}>-\eps_0 \sqrt{v_0R-g_0 p^0}
\ee
which is satisfied automatically if $\eps_0=1$ whereas if $\eps_0=-1$ it requires $g_0p^0>0$. More generally $r_h>r_i$ is a rather complicated expression which puts bounds on the allowed charges. It is difficult to analyze in complete generality but manageable in any given example.

Assuming that these conditions are satisfied we can use the results of \cite{Halmagyi:2013qoa} to analyze regular horizon geometries. Consider a black hole configuration with charges $(p^0,q_i)$ and gauging parameters $(g^i,g_0)$, in homogeneous $d$-geometries. The attractor equations give an expression of the horizon radius in terms of the charges and the $d$-tensor as derived in \cite{Halmagyi:2013qoa}, which reads
\begin{eqnarray}
R_2^4 &=& \frac{-a_4 \pm \sqrt{a_4^2-4 a_0 a_8}}{2a_8} \label{radius}\\
a_0 &=& -\frac{1}{16} p^0 \hd^{ijk} \nonumber\\
a_4 &=& \frac{9}{16} d_{g,i} \hd^{ilm} q_l q_m - (p^0g_0 + g^i q_i)^2 \nonumber\\
a_8 &=& 4 g_0 d_g= (\ell_{AdS})^{-4}>0 \nonumber\\
p^0 g_0&-&q_i g^i =1\,.
\end{eqnarray}
Still, a necessary condition for the existence of the horizon is that $R_2^2>0$. Notice that the sign of $g_0$ is chosen accordingly to the condition that $a_8>0$. If we use the last constraint on the charges, we can write 
\begin{eqnarray}
-4a_0a_8&=& d_{g}\, \hd_{q} (1+q_i g^i)\ ,\nonumber\\
a_4&=&\frac{9}{16}d_{g,i}\hd^i_q-(1+2q_i g^i)^2\ .
\end{eqnarray}
Notice that if $-4a_0a_8>0$ there is always a choice of sign in \eqref{radius} for which the radius is positive. But the sign of $-4a_0a_8$ can be driven to positive or negative by the choice of charges $q_i$. In that case, whatever the sign of $a_4$, there will always be a solution of the attractor equations for which the radius is positive. 
Indeed, let us consider a rescaling of all the $q$'s charges of the solution by a factor $\alpha\gtrless0$. This leads to 
\begin{eqnarray}
(-4a_0a_8)_{\alpha}&=& \alpha^3 d_g\, \hd_q (1+\alpha q_i g^i)\ ,\nonumber\\
(a_4)_{\alpha}&=&\alpha^2\frac{9}{16}d_{g,i}\,\hd^i_g-(1+2\alpha g_iq^i)^2\ .
\end{eqnarray}
Independently on what is the sign of $(a_4)_{\alpha}$, then, we can chose a small enough $\alpha\in \mathbb R$ for which $1+\alpha q_i g^i>0$. Then, depending on the sign of $\hd_q$, we can fix the sign of $(-4a_0a_8)_{\alpha}$ to be positive by requiring $\alpha\lessgtr 0$. This is enough to ensure that there is a root in the attractor equation corresponding to a positive $R_2^4$. 

\section{Conclusions}

In this work we have studied four dimensional $\cN=2$ FI-gauged supergravity theories where the scalar manifold is a homogeneous very special K\"ahler geometry. In these models we have found quarter-BPS black hole solutions with vanishing axions and constant phase $\psi$.

There are numerous interesting outstanding questions regarding this variety of AdS black holes. In the work \cite{Halmagyi:2013qoa} we considered models with arbitrary gauging parameters and arbitrary dyonic charges. It was found that that solution space of supersymmetric horizon geometries has real dimension $2n_v$; there are $2n_v+2$ charges and two constraints. It seems to be a well posed and reasonable open problem to solve for the most general quarter-BPS black hole in these theories with complex scalar fields and answer in the affirmative or otherwise whether every supersymmetric horizon geometry can be completed to a UV AdS$_4$ solution. This is very difficult to attack numerically but should one obtain the general analytic result, it would seem to be a reasonable question.
 
A key step in pursuing such an objective is a better understanding of black holes where the phase of the supersymmetry parameter is non-constant. In the current work we have only found black holes with constant phase but the full space of static BPS, AdS$_4$ black holes will surely include those with non-constant spinors. This could be quite challenging, for example with hypermultiplets all known solutions have non-trivial axions but there are certain solutions with  constant $\psi$ \cite{Donos:2012sy, Halmagyi:2013sla} while the general solution has varying $\psi$ \cite{Halmagyi:2013sla} and then analysis is significantly more complicated. We have argued  that all black holes with trivial axions will satisfy either \eq{gauge1} or \eq{gauge2} and thus have constant $\psi$ but have not found a proof of this statement.

A more modest objective could hopefully be realized using just the results of the current work. That is to determine whether every horizon geometry from \cite{Halmagyi:2013qoa} with vanishing axions and which satisfies \eq{gauge1} arises as the IR of the black holes in section \ref{sec:BHspecial}. 

At least to these humble authors, we find the origin of the ansatz \eq{v0ansatz}-\eq{Miansatz} and the ansatz in \cite{Cacciatori:2009iz} to be fairly mysterious. We have shown that it works just fine but we would certainly be comforted to have a deeper understanding of {\it why} it works. It is natural to speculate that a dimensional reduction to three dimensions \cite{Breitenlohner:1987dg} could aid this understanding, since such a reduction clarifies various issues for ungauged supergravity black holes \cite{Gaiotto2007e}. Another challenging approach would be to explicitly integrate the BPS equations rather than making the ansatz \eq{v0ansatz}-\eq{Miansatz}.

Hopefully these results will be a few steps along the road to a complete solution of supersymmetric static black holes in four dimensional gauged supergravity.

\vskip 1cm
\noindent {\bf Acknowledgements}
We would like to thank Guillaume Bossard,  Sergio Ferrara and Alessio Marrani for enjoyable and useful discussions. A.G. acknowledges support by the Netherlands Organization for Scientic Research (NWO) under the VICI grant 680-47-603

\begin{appendix}
\section{Special Geometry Conventions}

This material is all standard but we include it to make our conventions clear and in particular to be straight with our numerical factors. The prepotential we use 
\bea
F&=& -d_{ijk} \frac{X^i X^j X^k}{X^0} \label{Prepotential}
\eea
and we use special co-ordinates
\bea
X^\Lam&=& \bpm1 \\ z^i\epm\,,\quad\quad z^i = x^i + i y^i\,. \label{XLamexplicit}
\eea
From this we obtain that the dual sections $F_\Lam=\del_\Lam F$ are
\bea
F_\Lam&=& \bpm d_{ijk} z^i z^j z^k \\ -3 d_{z,i} \epm \label{FLamexplicit}
\eea
and the K\"ahler potential is
\bea
e^{-K}&=& 8  d_{ijk} y^i y^j y^k
\eea
so that the moduli space is constrained by $y^{i}>0$, $i=1,2,3$.
The symplectic form is given by
\bea
\Om&=& \bpm 0 & -1\!\!1 \\ 1\!\!1 & 0\epm\,.
\eea

We use the following shorthand for contraction with the symmetric tensors $d_{ijk}$ and $\hd^{ijk}$ of any component $g_i$ and $g^i$ taken from the matter couplings in any symplectic vector $(g^0,g^i,g_0,g_i)$:
\bea
&&d_g=d_{ijk} g^i g^j g^k\,,\ \ \ d_{g,i}= d_{ijk} g^j g^k\,,\ \ \ \ d_{g,ij}=d_{ijk} g^k\,, \non \\
&&\hd_g=\hd^{ijk} g_i g_j g_k\,,\ \ \ \hd_{g}^{i}= \hd^{ijk} g_j g_k\,,\ \ \ \ \hd_g^{ij}=\hd^{ijk} g_k\,.
\eea
For homogeneous spaces the matrix  $d_{g,ij}$ is invertible. The results of this paper do not need the explicit form of its inverse: we simply write $(d_g^{-1})^{ij}$ for the matrix that satisfies
\be
(d_g^{-1})^{ij} d_{g,jk}=\delta^i_k\,.
\ee
The metric on $\cM_v$ is given by
\bea
g_{ij} &=& -\frac{3}{2} \frac{d_{y,ij}}{d_y}+\frac{9}{4} \frac{d_{y,i}d_{y,j}}{d_y^2} 
\eea
and its inverse by
\bea
g^{ij}&=& -\frac{2}{3}d_y (d_y^{-1})^{ij}+2y^i y^j\,.
\eea
The following matrix is used in the presentation of the BPS equations
\bea
\cM=\bpm 1 & -\cR \\ 0 & 1 \epm. \bpm \cI & 0 \\ 0 & \cI^{-1} \epm. \bpm 1 & 0 \\ -\cR & 1 \epm=\bpm A & B \\ C & D \epm  \label{Mdef}
\eea
with
\bea
A&=& \cI +\cR \cI^{-1} \cR\,,\quad\quad  D= \cI^{-1}\,,\quad\quad B=C^T =-\cR \cI^{-1}
\eea
and where
\be
\cN_{\Lam\Sig}=\cR_{\Lam\Sig}+ i\, \cI_{\Lam\Sig}
\ee
is the symplectic matrix such that 
\be
M_\Lam = \cN_{\Lam\Sig} L^\Sig\,.
\ee
In addition $(\cR,\cI)$ give the vector kinetic and topological terms in the Lagrangian \eq{N2action}.
One can quite easily check that $\cM$ satisfies the identity
\be
\Om\cM \cV=-i \cV\,.
\ee
\section{Homogeneous Very Special K\"ahler Spaces} \label{app:homo}
For a homogeneous very special K\"ahler geometry we have the constant tensor
\bea
\hd^{ijk} =\frac{g^{il}g^{jm} g^{kn} d_{ijk}}{d_y^2}
\eea
which satisfies the relations
\bea
\hd^{ijk}d_{jl(m} d_{np)k} &=&\frac{16}{27} \Bslb  \delta^i_{l}d_{mnp} + 3 \delta^i_{(m} d_{np)l}  \Bsrb\,, \\
d_{ijk}d^{jl(m} d^{np)k} &=&\frac{16}{27} \Bslb  \delta_i^{l}\hd^{mnp} + 3 \delta_i^{(m} \hd^{np)l}  \Bsrb\,.
\eea
These in turn imply
\bea
\hd^{ijk}d_{j(lm} d_{np)k} &=&\frac{64}{27} \delta^i_{(l}d_{mnp)}\,, \label{dhdd}\\
d_{ijk}\hd^{j(lm} \hd^{np)k} &=&\frac{64}{27} \delta_i^{(l}\hd^{mnp)} \label{ddhdh} \,.
\eea
Using this we can solve the following equation which often appears in our work
\bea
F^i &=&\hd^{ijk} G_j G_k\ \ \ \Rightarrow\ \ \ G_i=\pm \sqrt{\frac{27}{64}} \frac{d_{ijk} F^j F^k}{\sqrt{d_F}}\,, \label{hdEq} \\
G_i &=&d_{ijk} F^j F^k\ \ \ \Rightarrow\ \ \ F^i=\pm \sqrt{\frac{27}{64}} \frac{\hd^{ijk} G_j G_k}{\sqrt{\hd_G}}\,.
\eea
One can also use $\hd^{ijk}$ to express the complex scalar fields in terms of the sections
\bea
z^i&=& \frac{3}{8}\frac{d_y}{d_z} \hd^{ijk} M_j M_k \label{yMM}
\eea
Other identities we find useful are
\bea
(d_g^{-1})^{ij}&=& \frac{1}{d_g}\Bslb \frac{27}{16} \hd^{ijk} d_{g,k}-3 g^i g^j \Bsrb\,, \\
(\hd_g^{-1})^{ij}&=& \frac{1}{\hd_g}\Bslb \frac{27}{16} d_{ijk} \hd_{g}^{k}-3 g_i g_j \Bsrb\,.
\eea

The quadratic and quartic invariants are given by
\bea
\cI_2 (a^\Lam ,b_\Lam)&=& -\frac{d_{ijk} a^i a^j a^k}{a^0}- \frac{1}{2} a^\Lam b_\Lam \label{quadratic}\\
\cI_4 (a^\Lam ,b_\Lam)&=& - \blp a^\Lam b_\Lam \brp^2+ \frac{1}{16} a^0 \hd^{ijk}  b_i b_j b_k- 4b_0 d_{ijk}  a^i a^j a^k +\frac{9}{16}d_{ijk} \hd^{ilm} a^j a^k b_l b_m \,. \label{quartic}
\eea

\end{appendix}


\providecommand{\href}[2]{#2}\begingroup\raggedright\endgroup

\end{document}